\documentclass[12pt]{article}
\RequirePackage[OT1]{fontenc}
\RequirePackage[authoryear]{natbib}

\usepackage{xcolor}
\usepackage{lineno,hyperref}
\usepackage{mathrsfs}
\usepackage{amstext,amsthm,amsmath,mathtools}
\usepackage{amssymb}
\usepackage{esint}
\usepackage{grffile} 
\usepackage{booktabs,multirow,dsfont}
\usepackage{algorithm}
\usepackage[noend]{algpseudocode}
\usepackage{varwidth}
\usepackage{subcaption}
\usepackage{siunitx}
\usepackage[english]{babel}
\usepackage{verbatim} 
\usepackage{import}
\usepackage{lscape}
\usepackage{color}

\def\Q{\mathbb{Q}}
\newtheorem{proposition}{Proposition}[section]
\newtheorem{assumption}{Assumption}[section]
\newtheorem{theorem}{Theorem}[section]
\newtheorem{remark}{Remark}[section]

\begin{document}

\title{\large\bf A New Nonparametric Estimate of the Risk-Neutral Density with Applications to Variance Swaps}
\author{Liyuan Jiang$^{1}$, Shuang Zhou$^{1}$, Keren Li$^{1}$, Fangfang Wang$^{2}$ and Jie Yang$^{1}$\\
		$^1$University of Illinois at Chicago and $^2$University of Wisconsin-Madison}
\maketitle

\begin{abstract}
We develop a new nonparametric approach for estimating the risk-neutral density of asset prices and reformulate its estimation into a double-constrained optimization problem. We evaluate our approach using the S\&P 500 market option prices from 1996 to 2015. A comprehensive cross-validation study shows that our approach outperforms the existing nonparametric quartic B-spline and cubic spline methods, as well as the parametric method based on the Normal Inverse Gaussian distribution. As an application, we use the proposed density estimator to price long-term variance swaps, and the model-implied prices match reasonably well with those of the variance future downloaded from the CBOE website.
\end{abstract}


\section{Introduction}

A financial derivative, such as option, swap, future, or a forward contract, is an asset that is contingent on an underlying asset. Its fair price can be obtained by calculating the expected future payoff under a risk-neutral probability distribution. Therefore, the  problem of pricing a derivative can be addressed via estimating the risk-neutral density of the future payoff of the underlying asset.  On the other hand, the market prices of the derivatives traded in a financial market reveal information about the risk-neutral density. \cite{breedenlit:tex} was among the first to use option prices to estimate the risk-neural probability distribution of the future payoff of the underlying asset. Among the financial products that can be used to recover the risk-neutral density, European options are the most common ones, which give the investors rights to trade assets at a pre-agreed price (i.e., strike price) at the maturity date. Among all the underlying assets that options are written on, Standard \& Poor's 500 Index (S\&P 500) is a popular one, which aggregates the values of stocks of 500 large companies traded on American stock exchanges and provides a credible view of American stock market for investors. 

There are a plethora of approaches towards recovering risk-neutral density functions in the literature (see, for example, \cite{robertnik:tex} for an extensive review). Parametric approaches typically specify a statistical model for the risk-neutral density and  the structural parameters are recovered by solving an optimization problem.  For instance, \cite{jarrowrudd:tex} used a lognormal distribution; \cite{melickthomas:tex} considered a mixture of lognormal distributions proposed by \cite{ritchey:tex};  \cite{sherrick1992:tex} employed a three-parameter Burr distribution, called the Burr family, which covers a broad range of shapes, including distributions similar to gamma, lognormal, and J-shaped beta. Another commonly used probability distribution in the literature of derivative pricing is the Generalized Hyperbolic distribution that contains Variance Gamma, Normal Inverse Gaussian, and $t$ distributions as special cases (see, for instance, \cite{eriksson2009normal} and \cite{eric:pctex}). 

Nonparametric procedures, by contrast, are free from distributional assumptions on the underlying asset and thus achieve more flexibility than parametric methods.  For example, \cite{anarehaluis:tex} used cubic spline functions to model the unknown risk-neutral density. An estimated density is numerically obtained by solving a quadratic programming problem with a convex objective function and non-negativity constraints. They deliberately chose more knots than option strikes for higher flexibility. \cite{shlee:tex} approximated the risk-neutral cumulative distribution function using Quartic B-splines with power tails and the minimum number of knots that meet zero bid-ask spread. Their estimation was based on out-of-the-money option prices.

In this paper, we propose a simpler but more powerful nonparametric solution using piecewise constant functions to estimate the risk-neutral density. It is easy to implement since the estimating problem is formulated as a weighted least squared procedure.  
It is more powerful since our method can recover the risk-neutral density more effectively with all available option market prices without screening. Furthermore, our solution provides a practical way to explore profitable investment opportunities in financial markets by comparing the estimated prices and the corresponding market prices.

The rest of this paper is structured as follows. Section~\ref{sec:method}  introduces the proposed nonparametric approach after reviewing cubic splines, Quartic B-splines, and the Normal Inverse Gaussian (NIG) parametric approaches in the literature. In Section~\ref{sec:example}, we run comprehensive cross-validation studies using real data to compare different methods. In Section~\ref{sec:swap}, we apply the proposed nonparametric approach to price variance swaps, which is challenging in practice \cite{zhulian:tex, prl:tex}. We conclude in Section~\ref{sec:conclude}. The proofs and more formulae are collected in the Appendix.

\section{Methodologies}\label{sec:method}

In this section, we first provide a brief review of the cubic splines, quartic B-splines, and NIG approaches in the literature for recovering the risk-neutral density (RND). Then we introduce the proposed piecewise constant (PC) nonparametric approach with least square and weighted least square procedures.

\subsection{Nonnegative cubic spline estimate for RND}\label{sec:cubic}

Given the current trading date $t$ and the expiration date $T$ of European options, let $[K_1, K_q]$ be the range of strike prices of all available options traded in the market. \cite{anarehaluis:tex} considered $s+1$ equally spaced knots for a cubic spline with $K_1 = x_1 < x_2 < x_3 < \cdots < x_s < x_{s+1} = K_q$. These knots are not necessarily a subset of the available strikes. Nevertheless, the closer these knots are to the strikes, the better their solution is. \cite{anarehaluis:tex} also claimed that the number of knots should not be very much larger than the number of distinct strikes.

For the sake of non-negativity of the estimated RND, \cite{anarehaluis:tex}'s solution is much more complicated and computationally expensive than the usual cubic spline estimates. For comparison purposes, we keep only the constraints that ensure the non-negativity of the density function on knots in their optimization procedure. By evaluating the difference between the estimated fair prices and the market prices of options, if our approach achieves higher accuracy than the cubic spline estimate with less constraints, then our approach is considered to be superior to that of \cite{anarehaluis:tex}.

When it comes to practical implementation, \cite{anarehaluis:tex} suggested eliminating option prices that led to potential arbitrage opportunities according to bid-ask interval, put-call parity, monotonicity, and strict convexity. They also generated ``fake" call option prices using put-call parity to eliminate ``artificial" arbitrage opportunities. Our comprehensive studies in Section~\ref{sec:example} show that their screening and cleaning procedure may result in substantial information loss.

\subsection{Quartic B-spline estimate}\label{sec:bspline} 

\cite{shlee:tex} adopted a uniform quartic B-spline to estimate the risk-neutral cumulative distribution function (CDF). They used power tails to extrapolate outside the strike price range. 
\cite{shlee:tex} suggested using only the out-of-the-money (hereafter OTM) options to estimate the CDF, including OTM call options whose strikes are higher than the underlying asset price, and OTM put options whose strikes are lower than the underlying asset price. 
OTM options are typically cheaper than in-the-money (ITM) options and are considered to be more liquid as well.
Nevertheless, our case studies in Section~\ref{sec:example} show that ITM options may help recover the risk-neutral distribution as well.

Due to fewer parameters, the quartic B-spline estimate is computationally more efficient than the nonnegative cubic spline approach. 
\cite{shlee:tex} chose the number of knots needed as the minimum number that satisfies zero bid-ask pricing spread. 
They also suggested eliminating options that violate monotonicity and strict convexity constraints.

\subsection{NIG parametric approach}\label{sec:nig}

For comparison purposes, we choose one parametric approach for approximating the risk-neutral density, as suggested by \cite{eriksson2004approximating} and \cite{eriksson2009normal}. It is based on the Normal Inverse Gaussian (NIG) distribution, which belongs to the Generalized Hyperbolic class and can be characterized by its first four moments, i.e., mean, variance, skewness, and kurtosis. According to \cite{bakshi:pctex}, these four moments can be estimated by the OTM European call and put options. One major issue with NIG density estimate is that, as shown in \cite{eric:pctex}, the feasibility of NIG approach drops down as the time to maturity increases, since more estimated skewness and kurtosis pairs fall outside the feasible domain of the NIG distribution.

\subsection{The proposed piecewise constant nonparametric approach}\label{sec:nonpara}

The piecewise constant (PC) approach that we propose in this paper is nonparametric by nature, and it is simpler but more efficient. Let $S_t$ and $S_T$ stand for the current price  of an equity on day $t$ and the future price on day $T$. To estimate the risk-neutral density function $f_{\Q}$ of $\log(S_T)$ conditional on the information up to day $t$, we propose to use a piecewise constant function, or a step function, to approximate $f_{\Q}$, with all distinct strike prices as knots.
The constants in the step function are estimated by solving an optimization problem subject to certain constraints. By forcing the constants to be nonnegative, the non-negativity of the estimated risk-neutral density is guaranteed.

To be precise, suppose that we have a collection of market prices of European put and call options that are traded on date $t$ and expire on date $T$. Let $\{K_1, K_2, \ldots, K_q\}$ represent the distinct strikes in ascending order, and $\mathcal{C}$ be the collection of indices for call options and $\mathcal{P}$ for put options.  
Then $\mathcal{C} \cup \mathcal{P} = \{  1, 2, \ldots, q \}$.  Let $m = |\mathcal{C}|$ and $n = |\mathcal{P}|$ be the numbers of calls and puts, respectively. Then $m+n \geq q$~. 

Given a risk-neutral density $f_{\Q}$, the fair prices of put option and call option with strike $K_i$ are 
\begin{eqnarray*} 
	P_i &=& \mathbb{E}_t^{\Q}e^{- R_{tT}} (K_i - S_T)_+ = e^{- R_{tT}} \int_{-\infty}^{\log K_i} (K_i - e^y) f_{\Q}(y) dy, \\
	C_i &=& \mathbb{E}_t^{\Q}e^{- R_{tT}} (S_T - K_i)_+ = e^{- R_{tT}} \int_{\log K_i}^{\infty}  (e^y - K_i) f_{\Q}(y) dy, 
\end{eqnarray*}
respectively, where $R_{tT}$ stands for the cumulative risk-free interest rate from $t$ to $T$; that is, $\$$1 on day $t$ ends for sure with $e^{R_{tT}}$ dollars on day $T$. We denote by $r_t$ the risk-free interest rate over the period $[t, t+1]$, which is obtained from risk-free zero-coupon bond, and clearly $R_{tT}=\sum_{j=t}^{T-1} r_j$~.

To account for the risk-neutral density outside the range $[K_1, K_q]$, we add $K_0=K_1/c_K$ and $K_{q+1}= c_K K_q$, where $c_K>1$  is a predetermined constant that can be chosen by means of cross-validation or prior knowledge (see details in Section~\ref{sec:example}).
We then use a piecewise constant function $f_\Delta$ to approximate $f_{\Q}$; that is,  
\begin{equation}\label{eqn:pdfpc}
f_\Delta(y) = a_l, \quad \text{ for } \log K_{l-1} < y \leq \log K_l, ~ l = 1, 2, \ldots, q+1, 
\end{equation}
and zero elsewhere. Here $\Delta = \{\log K_1, \ldots, \log K_q\}$ stands for the collection of distinct strikes in log scale, and $\{a_l, l=1, \ldots, q+1 \}$ are nonnegative constants satisfying 
\begin{equation} \label{eq:unityct}
\sum_{l=1}^{q+1} a_l \log \frac{K_l}{K_{l-1}} =1
\end{equation} 
due to the condition $\int_{-\infty}^{+\infty} f_\Delta(y) dy =1$. 

Given the approximate risk-neutral density $f_\Delta$, the estimated put and call  prices  with strike $K_i$ are
\begin{eqnarray} 
\hat{P}_i &=& e^{- R_{tT}} \int_{-\infty}^{\log K_i} (K_i - e^y) f_\Delta(y) dy, \label{eq: estrnd1}\\
\hat{C}_i &=& e^{- R_{tT}} \int_{\log K_i}^{\infty}  (e^y - K_i) f_\Delta (y) dy,  \label{eq: estrnd2}
\end{eqnarray}
respectively, which are essentially linear functions of $a_1, \ldots, a_q$~.

\begin{proposition}\label{prop:callputs}
	Given $a_l\geq 0, l=1, \ldots, q+1$ satisfying \eqref{eq:unityct}, the estimated prices for put and call options with strike $K_i$ satisfy
	\begin{eqnarray}
	e^{R_{tT}} \hat{P}_i  &=& a_1 X^{(P)}_{i,1} +  \cdots + a_q X^{(P)}_{i,q} +  X^{(P)}_{i,q+1}, \label{eq:equ3} \\
	e^{R_{tT}} \hat{C}_i &=& a_1 X^{(C)}_{i,1} + \cdots + a_q X^{(C)}_{i,q} +  X^{(C)}_{i,q+1}, \label{eq:equ4}
	\end{eqnarray}
	where $X^{(P)}_{i,l}  = X^{(p)}_{i,l} -\log (K_l / K_{l-1}) (\log c_K)^{-1} X^{(p)}_{i,q+1}$, $X^{(C)}_{i,l} = X^{(c)}_{i,l} - \log (K_l / K_{l-1})$ $(\log c_K)^{-1}$ $X^{(c)}_{i,q+1}$, $l = 1, 2, \ldots, q$; $X^{(P)}_{i,q+1}  = X^{(p)}_{i,q+1} (\log c_K)^{-1}$, $X^{(C)}_{i,q+1} = X^{(c)}_{i,q+1}$ $(\log c_K)^{-1}$; and $X^{(p)}_{i,l} = [K_i  \log(K_l / K_{l-1}) - (K_l - K_{l-1})]   \cdot  \mathds{1} (K_i \geq K_l )$, $X^{(c)}_{i,l} = [ (K_{l} - K_{l-1}) - K_i \log (K_{l}/K_{l-1})] \cdot  \mathds{1} (K_i < K_{l})$, $l = 1, 2, \ldots, q+1$.
\end{proposition}
\noindent The proof of Proposition~\ref{prop:callputs} is relegated into Appendix~\ref{sec:proofcallputs}. 

The unknown parameters $a_1, \ldots, a_{q+1}$ are estimated by minimizing the following least square (LS) objective function 
\begin{equation} \label{eq:objfunc}
L(a_1, \ldots, a_{q+1})  = \frac{1}{m+n}\left[ \sum_{i \in \mathcal{C}}(\hat{C}_i - \tilde{C}_i)^2 + \sum_{i  \in \mathcal{P} } (\hat{P}_i - \tilde{P}_i)^2  \right] 
\end{equation} 
subject to  $a_l \geq 0$, $l=1,2,\ldots, q+1$, and Equation~\eqref{eq:unityct}, where $\tilde{C}_i$ and $\tilde{P}_i$ are market prices of call option and put option, respectively, with strike $K_i$~. If there exists a risk-neutral density $f_{\Q}$, we have $C_i = \tilde{C}_i, i \in \mathcal{C}$ and $P_i = \tilde{P}_i, i \in \mathcal{P}$. That is, the market prices are fair if there is no arbitrage in the financial market.

From an investment point of view, because a more expensive option tends to be less liquid, an alternative approach to determining $a_1, \ldots, a_{q+1}$ is to minimize a weighted least square (WLS) objective function
\begin{equation}\label{eq:wlsobj}
W(a_1, \ldots, a_{q+1})  = \frac{1}{m+n}\left[ \sum_{i \in \mathcal{C}}\left(\frac{\hat{C}_i - \tilde{C}_i}{\tilde{C}_i}\right)^2 + \sum_{i  \in \mathcal{P} } \left(\frac{\hat{P}_i - \tilde{P}_i}{\tilde{P}_i}\right)^2  \right]. 
\end{equation}
The WLS estimate is in favor of OTM options over ITM options, in that  OTM options are typically less expensive and more liquid.

\section{Pricing European Options}\label{sec:example}

In this section, we use the S\&P 500 European options to evaluate the performances of various RND estimators.

\subsection{S\&P 500 European option data}\label{sec:data1}

We consider European calls and puts written on the S\&P 500 indices from January 2, 1996 to August 31, 2015 in the US. The expiration dates are the third Saturday of the delivery month. 
Following \cite{prl:tex}, we keep only the options with positive bid prices, positive volumes, and with expiration more than seven days in our analysis. Similar to \cite{eric:pctex}, we categorize options into seven groups with expiration in $7 \sim 14$ days, $17 \sim 31$ days, $81 \sim 94$ days, $171 \sim 199$ days, $337 \sim 393$ days, $502 \sim 592$ days, and $670 \sim 790$ days, respectively, for the purpose of examining the effects of the length of maturity on pricing. The numbers of options and $(t, T)$ pairs under consideration are presented in Table~\ref{table:numofcp}.

\begin{table}
	\begin{center}
		\caption{Numbers of calls, puts, and $(t, T)$ pairs in different time-to-maturity categories (number of days to expiration)}
		\label{table:numofcp}
		\footnotesize
		\begin{tabular}{|l|r|r|r|r|r|r|r|}
			\hline
			\#Day & 7$\sim$14  & 17$\sim$31 & 81$\sim$ 94  & 171$\sim$199 & 337$\sim$393 & 502$\sim$592 & 670$\sim$790 \\ \hline
			\#Call & 72535 & 136019 & 34764  & 17367 & 13465   & 7985   & 5869   \\ \hline
			\#Put & 112862 & 205863 & 53648  & 27906  & 18982   & 14535   & 10104   \\ \hline
			\#$(t, T)$ & 2411 & 4206 & 2548  & 2306  & 2747   & 2536 & 1739   \\ \hline
		\end{tabular}
		\normalsize
	\end{center}
\end{table}

\subsection{Comprehensive comparisons with existing methods}

We use the S\&P 500 European options to evaluate the performances of the following methods: the parametric NIG estimate, quartic B-spline (Bspline) estimate, the nonnegative cubic spline estimates with either least square criterion (Cubic + LS) or weighted least square criterion (Cubic + WLS), as well as the proposed piecewise constant estimate with either least square or weighted least square objective function using OTM options only (PC + LS + OTM or PC + WLS + OTM) or using all available options (PC + LS + ALL or PC + WLS + ALL). All the comparisons are made based on their ability of recovering option market prices.

For each of the seven time-to-maturity categories listed in Table~\ref{table:numofcp}, we randomly select 200 pairs of $(t, T)$. For each pair, the market prices of calls and puts are collected. The aforementioned approaches are applied to estimate the RND of the underlying asset at time $T$. We then use the estimated RND to obtain $\hat{C}_i$ and $\hat{P}_i$~. Discrepancy between the market prices and the estimated prices is assessed by means of the absolute error $L_a$ and the relative error $L_r$ defined as below
\begin{eqnarray*}
	L_a^2  &=& \frac{1}{|\mathcal{C}_t| + |\mathcal{P}_t|}\left[\sum_{i\in \mathcal{C}_t} (\hat{C}_i - \tilde{C}_i)^2 + \sum_{i\in \mathcal{P}_t} (\hat{P}_i - \tilde{P}_i)^2\right], \\
	L_r^2 &=& \frac{1}{|\mathcal{C}_t| + |\mathcal{P}_t|}\left[\sum_{i\in \mathcal{C}_t} (\hat{C}_i/ \tilde{C}_i -1)^2 + \sum_{i\in \mathcal{P}_t} (\hat{P}_i/\tilde{P}_i-1)^2\right],
\end{eqnarray*}
where $\mathcal{C}_t$ (or $\mathcal{P}_t$) refers to the collection of indices of call (or put) options used for testing purposes. In Table~\ref{my-label} and Table~\ref{my-label2}, we choose $\mathcal{C}_t$ and $\mathcal{P}_t$ to be either all available OTM options or ITM options. We report the average $L_a$ and $L_r$ over the 200 randomly selected pairs of $(t, T)$ for each estimation approach. The columns labeled ``200'' show the actual number of pairs that  yield a valid RND estimate. The higher the count, the more effective the method is. As explained in Section~\ref{sec:nig}, the NIG approach is very picky in selecting calls and puts. 
For B-spline and Cubic methods, following the same filtering procedures as in  \cite{anarehaluis:tex} and \cite{shlee:tex}, we observe that fewer options become available as the time-to-maturity increases, which results in substantial information loss. On the contrary, our PC methods with LS or WLS are feasible for almost all cases, especially when using both ITM and OTM options.

In terms of the absolute error $L_a$ and the relative error $L_r$ computed for different combinations of time-to-maturities and RND estimates, our PC estimates are more stable and accurate than the other three  approaches. As illustrated in Tables~\ref{my-label} and \ref{my-label2}, the proposed PC methods always yield the lowest $L_a$ or $L_r$, regardless of the type of options used. 
In order for a cross-sectional comparison among all the approaches, only OTM options are considered when using the proposed PC approach to price options (i.e., PC + LS + OTM or PC + WLS + OTM).  
But as far as practical implementation is concerned, we would recommend using all available option prices, including both ITM and OTM options. In particular, if the goal is to obtain the most precise price, we recommend ``PC+LS+ALL'', in that it controls absolute error $L_a$ the best; if one seeks for higher return on investment, we would recommend ``PC+WLS+ALL'' instead, which controls relative error $L_r$ the best.

\begin{table}[]
	\centering
	\caption{Comprehensive Comparison of Different RND Estimates - Part I}
	\label{my-label}
	\centering\small\setlength\tabcolsep{2pt}
	\footnotesize
	\hspace*{-1cm}\begin{tabular}{|l|c|r|r|r|r|r|r|r|r|r|r|r|r|} 
		\hline
		\multicolumn{2}{|r|}{Time-to-maturity}     & \multicolumn{3}{c|}{7$\sim$14} & \multicolumn{3}{c|}{17$\sim$31} & \multicolumn{3}{c|}{81$\sim$94} & \multicolumn{3}{c|}{171$\sim$199}  \\ \hline
		Method & Test  & $L_a$       & $L_r$       & 200  & $L_a$        & $L_r$       & 200  & $L_a$        & $L_r$       & 200  & $L_a$          & $L_r$         & 200 \\ \hline
		NIG         & ITM & 1.823   & 0.058   & 145  & 2.293    & 0.057   & 110  & 5.902    & 0.095   & 91   & 14.344     & 0.158     & 143 \\ \hline
		& OTM & 0.772   & 0.569   & 145  & 1.669    & 0.533   & 110  & 5.404    & 0.769   & 92   & 10.445     & 0.771     & 143 \\ \hline
		B-spline     & ITM & 27.031  & 0.107   & 133  & 30.404   & 0.140   & 156  & 33.019   & 0.124   & 93   & 23.086     & 0.144     & 30  \\ \hline
		& OTM & 1.638   & 15.102  & 133  & 9.981    & 64.444  & 156  & 5.037    & 7.153   & 93   & 11.783     & 12.655    & 30  \\ \hline
		Cubic + LS   & ITM & 3.645   & 0.218   & 102  & 1.055    & 0.028   & 76   & 2.254    & 0.041   & 77   & 69861.9  & 734.427   & 69  \\ \hline
		& OTM & 3.105   & 4.452   & 102  & 0.387    & 0.600   & 76   & 1.094    & 1.276   & 77   & 224532.5 & 15259.638 & 69  \\ \hline
		Cubic + WLS  & ITM & 4.696   & 0.236   & 102  & 1.286    & 0.034   & 76   & 2.977    & 0.049   & 77   & 66506.6  & 699.154   & 69  \\ \hline
		& OTM & 3.480   & 5.032   & 102  & 0.446    & 0.656   & 76   & 1.297    & 1.084   & 77   & 214119.1 & 14806.153 & 69  \\ \hline
		PC+LS+ALL  & ITM & {\bf 0.138}   & {\bf 0.005}   & 200  & {\bf 0.150}    & {\bf 0.004}   & 200  & {\bf 0.269}    & {\bf 0.004}   & 200  & {\bf 0.430}      & {\bf 0.004}     & 200 \\ \hline
		& OTM & 0.083   & 0.157   & 200  & 0.097    & 0.114   & 200  & 0.162    & 0.077   & 200  & 0.420      & 0.056     & 200 \\ \hline
		PC+WLS+ALL & ITM & 0.219   & 0.007   & 200  & 0.231    & 0.005   & 200  & 0.462    & 0.006   & 200  & 1.628      & 0.008     & 200 \\ \hline
		& OTM & 0.077   & {\bf 0.074}   & 200  & 0.090    & {\bf 0.064}   & 200  & 0.166    & 0.034   & 200  & 0.370      & 0.028     & 200 \\ \hline
		PC+LS+OTM  & ITM & 0.679   & 0.023   & 200  & 0.646    & 0.015   & 200  & 6.570    & 0.098   & 198  & 25.042     & 0.171     & 198 \\ \hline
		& OTM & {\bf 0.053}   & 0.098   & 200  & {\bf 0.074}    & 0.086   & 200  & {\bf 0.153}    & 0.043   & 198  & {\bf 0.275}      & 0.036     & 198 \\ \hline
		PC+WLS+OTM & ITM & 0.913   & 0.029   & 200  & 0.803    & 0.019   & 200  & 9.073    & 0.114   & 198  & 25.135     & 0.172     & 198 \\ \hline
		& OTM & 0.121   & 0.077   & 200  & 0.121    & 0.065   & 200  & 0.308    & {\bf 0.034}   & 198  & 0.364      & {\bf 0.025}     & 198 \\ \hline
	\end{tabular}
	\normalsize
\end{table}

\begin{table}[]
	\centering
	\caption{Comprehensive Comparison of Different RND Estimates - Part II}
	\label{my-label2}
	\centering\small\setlength\tabcolsep{2pt}
	\footnotesize
	\hspace*{-1cm}\begin{tabular}{|l|c|r|r|r|r|r|r|r|r|r|}
		\hline
		\multicolumn{2}{|r|}{Time-to-maturity}     & \multicolumn{3}{c|}{337$\sim$393} & \multicolumn{3}{c|}{502$\sim$592}  & \multicolumn{3}{c|}{670$\sim$790} \\ \hline
		Method & Test  & $L_a$            & $L_r$        & 200 & $L_a$          & $L_r$         & 200 & $L_a$          & $L_r$        & 200 \\ \hline
		NIG         & ITM & 23.238     & 0.165    & 51  & 37.368     & 0.216     & 27  & 49.308     & 0.180    & 29  \\ \hline
		& OTM & 17.781     & 0.790    & 53  & 28.575     & 1.236     & 27  & 33.174     & 5.329    & 29  \\ \hline
		B-spline     & ITM & 146.941    & 0.255    & 4   & NA          & NA         & 0   & NA          & NA        & 0   \\ \hline
		& OTM & 146.941    & 0.255    & 4   & NA          & NA         & 0   & NA          & NA        & 0   \\ \hline
		Cubic + LS   & ITM & 251615.4 & 876.7  & 68  & 248235.4 & 778.5   & 75  & 47327.8  & 95.317   & 54  \\ \hline
		& OTM & 110639.6 & 1553.4 & 68  & 303111.7 & 24539.1 & 75  & 24203.2  & 2351.626 & 54  \\ \hline
		Cubic + WLS  & ITM & 250487.1 & 872.8  & 68  & 406119.5 & 1259.7  & 75  & 47364.3  & 95.391   & 54  \\ \hline
		& OTM & 110189.3 & 1547.4 & 68  & 517077.6 & 35028.3 & 75  & 24205.7  & 2353.728 & 54  \\ \hline
		PC+LS+ALL  & ITM & {\bf 4.907}      & {\bf 0.033}    & 200 & {\bf 7.406}      & 0.089     & 200 & 7.501      & 0.062    & 200 \\ \hline
		& OTM & 1.323      & 0.066    & 200 & 3.100      & 0.154     & 200 & 5.484      & 0.098    & 200 \\ \hline
		PC+WLS+ALL & ITM & 7.148      & 0.035    & 200 & 7.778      & {\bf 0.070}     & 200 & {\bf 6.556}      & {\bf 0.051}    & 200 \\ \hline
		& OTM & 2.256      & 0.028    & 200 & 4.382      & 0.044     & 200 & 6.268      & 0.048    & 200 \\ \hline
		PC+LS+OTM  & ITM & 79.914     & 0.320    & 192 & 92.636     & 0.439     & 197 & 86.013     & 0.404    & 194 \\ \hline
		& OTM & {\bf 1.150}      & 0.032    & 192 & {\bf 1.401}      & 0.050     & 197 & {\bf 1.153}      & 0.044    & 194 \\ \hline
		PC+WLS+OTM & ITM & 79.688     & 0.318    & 192 & 93.565     & 0.438     & 197 & 85.833     & 0.403    & 194 \\ \hline
		& OTM & 1.461      & {\bf 0.021}    & 192 & 2.1977      & {\bf 0.021}     & 197 & 1.479      & {\bf 0.018}    & 194 \\ \hline
	\end{tabular}
	\normalsize
\end{table}

\subsection{Consistency of PC estimates for fair prices}

Given distinct strike prices $K_1 < K_2 < \cdots < K_q$, the associated market prices of calls and puts, $\{\tilde{C}_i, i \in \mathcal{C}\}$ and $\{\tilde{P}_i, i \in \mathcal{P}\}$ respectively, traded on date $t$ with expiration date $T$ satisfy 
\begin{equation}\label{eq:CP_fQ} 
\tilde{C}_i = e^{- R_{tT}} \int_{\log K_i}^{\infty}  (e^y - K_i) f_{\Q}(y) dy,\>\>\> 
\tilde{P}_i = e^{- R_{tT}} \int_{-\infty}^{\log K_i} (K_i - e^y) f_{\Q}(y) dy,
\end{equation}
provided that a risk-neutral density $f_\Q$ of $\log S_T$ exists. That is, the market prices $(\tilde{C}_i, \tilde{P}_i)$ agree with the fair prices $(C_i, P_i)$~.

Recall that the proposed PC approach provides an approximation 
\begin{equation}\label{eq:fdelta}
f_\Delta (x) = \sum_{l=1}^{q+1} a_l {\mathbf 1}_{(\log K_{l-1},\ \log K_l]}(x)
\end{equation} 
to the RND $f_\Q$ where $(a_1, \ldots, a_{q+1})$ are such that minimize the absolute error $L(a_1, \ldots, a_{q+1})$ or the relative error $W(a_1, \ldots, a_{q+1})$. The estimated fair prices, $(\hat{P}_i, \hat{C}_i)$, calibrated using $f_\Delta$ are determined by Equations \eqref{eq: estrnd1} and \eqref{eq: estrnd2}.

Because $f_\Q$ is practically not unique, instead of measuring the distance between $f_\Delta$ and $f_\Q$, we would like to ask whether the prices obtained using $f_\Delta$ could recover the market prices well. The extensive numerical studies reported in Tables~\ref{my-label} and \ref{my-label2} corroborate this claim. This is further justified by the following theorem. 

\begin{theorem}\label{thm:consistency}
Suppose there exits a continuous risk-neutral density $f_\Q$ of $\log S_T$ satisfying $\int_0^\infty e^x f_\Q(x) dx < \infty$. Let $\Delta=\{\log K_1, \ldots, \log K_q\}$ be the collection of distinct strike prices in log scale with both call and put option market prices available. Then as $K_1 \rightarrow 0$, $K_q \rightarrow \infty$, $q\rightarrow \infty$, and $|\Delta|:=\max_{1\leq i < q} \log (K_{i+1}/K_i) \rightarrow 0$, we have
\[\frac{1}{2q}\left[\sum_{i=1}^q (\hat{C}_i - \tilde{C}_i)^2 + \sum_{i=1}^q (\hat{P}_i-\tilde{P}_i)^2\right]\longrightarrow 0.\]
\end{theorem}

\begin{remark}{\rm
Since $\tilde{C}_i = e^{- R_{tT}} \int_{\log K_i}^{\infty}  e^y f_{\Q}(y) dy - K_i e^{- R_{tT}} \int_{\log K_i}^{\infty}   f_{\Q}(y) dy$, 
then the condition $\int_0^\infty e^x f_\Q(x)$ $dx < \infty$ in Theorem~\ref{thm:consistency} is  necessary and sufficient for $\tilde{C}_i < \infty$.
}\end{remark}

\begin{remark}{\rm
		The proof for Theorem~\ref{thm:consistency} is relegated to Appendix~\ref{sec:theorem3.1proof}. It shows the existence of $(a_1, \ldots, a_{q+1})$ such that $\max_{1\leq i\leq q} |\hat{C}_i - \tilde{C}_i| < \epsilon$ and  $\max_{1\leq i\leq q} |\hat{P}_i - \tilde{P}_i| < \epsilon$ for any given $\epsilon > 0$ when $K_1, |\Delta|$ are sufficiently small and $K_q, q$ are sufficiently large. In other words, $|\hat{C}_i - \tilde{C}_i|, |\hat{P}_i - \tilde{P}_i|, i=1, \ldots, q$, can be uniformly small.		
}\end{remark}

\subsection{Detecting profitable opportunities}

Theorem~\ref{thm:consistency} provides analytical foundations for the consistency of the proposed PC method under the assumption of the existence of a continuous risk-neutral density. Nevertheless, the PC method is still applicable even when there is an arbitrage opportunity in the market. In this case, a significant difference between market price and its estimated fair price would be expected.

With a given set of market option prices, our nonparametric method can recover a fair option price for any strike price. From an investment point of view, we are able to detect options on the markets that are under or over priced.  It may not be adequate to claim arbitrage opportunities due to the lack of guarantee to earn and since there is a mature market system designed to catch such kind of difference among the option prices. Nevertheless, we can still report profitable investment opportunities for investors. 

\begin{figure}
	\centering
	\includegraphics[height = 6cm,width=12cm]{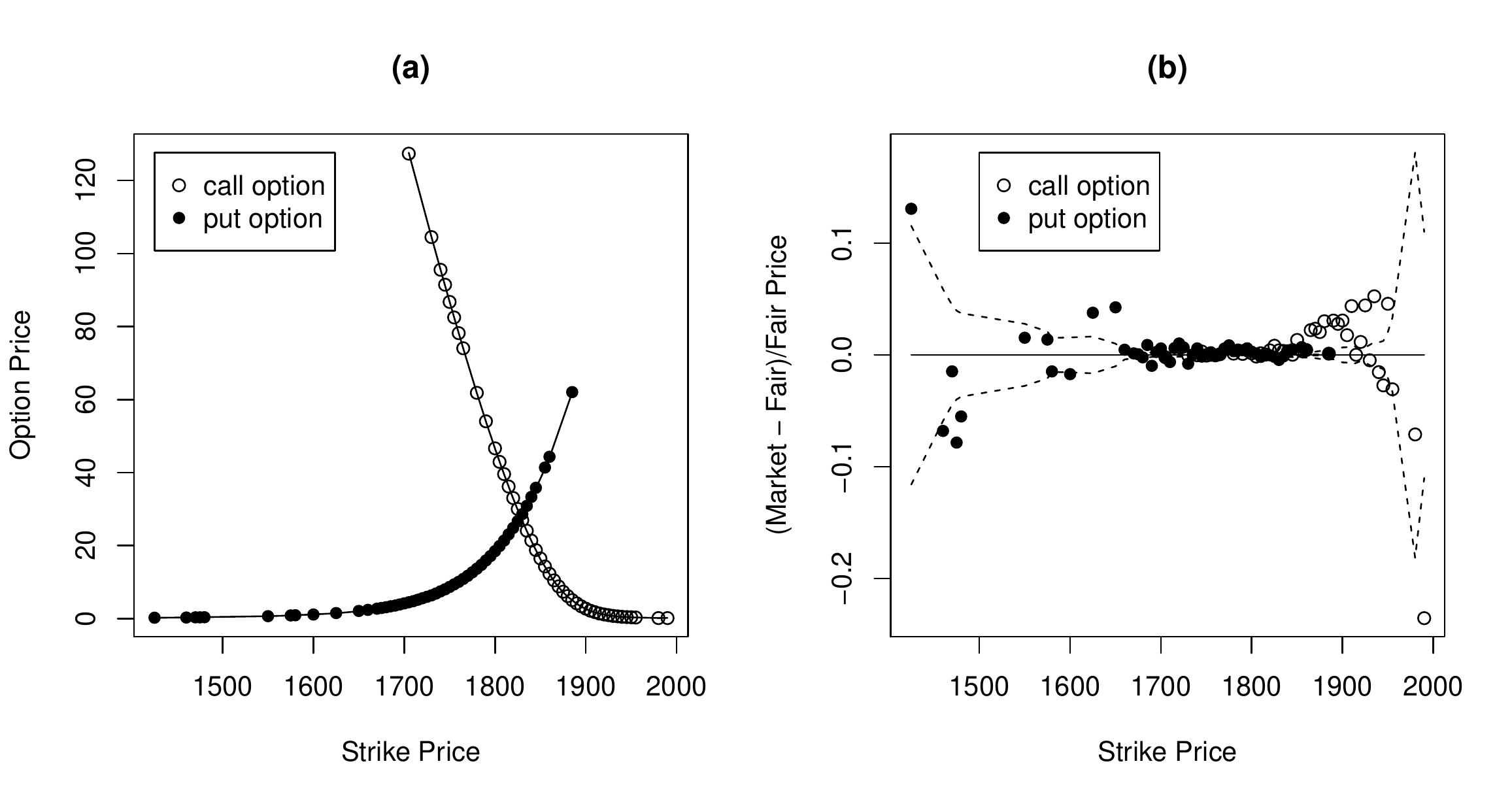}
	\caption{Leave-one-out cross validation with options traded on 04/14/2014 with expiration 5/9/2014 (round dot: market price; solid line: fair price based on PC; dash line: 95\% confidence interval based on bootstrap)}
	\label{fig:applicationinvest}
\end{figure}

In Figure~\ref{fig:applicationinvest}, we provide an illustrative example using $m+n=95$ available market prices of options traded on 04/14/2014 with expiration 5/9/2014. For each of the 95 options, we obtain its fair price based our PC+LS method using the market prices of the rest $m+n-1=94$ options. Then we compare the market price and the leave-one-out fair price, known as leave-one-out cross validation. Figure~\ref{fig:applicationinvest}(a) depicts $m+n=95$  market prices in dots and leave-one-out fair prices in solid line against the corresponding strike prices. It seems that they match each other very well.

To have a closer look at the difference between market price and fair price, we plot the relative difference, that is, (market price - fair price)/fair price, against strike price in Figure~\ref{fig:applicationinvest}(b).  The sign of the relative difference tells us whether the option is under or over priced. In addition, in order to check if the difference between a market price and its fair price is statistically significant, we bootstrap the rest of market prices 50 times to obtain a $95\%$ confidence interval of the fair price. The dash lines in Figure~\ref{fig:applicationinvest}(b) show the upper and lower ends of the bootstrap confidence intervals. When a market price falls outside its bootstrap confidence interval, we may report to investors that the corresponding option is significantly under/over priced compared with the market prices of the other options.

\section{Pricing Variance Swap}\label{sec:swap}

With an estimated risk-neutral density, one can calculate the fair price of any derivative whose payoff is a function of $S_T$. In this section, we apply the proposed method to price variance swaps. Our study shows that our fair prices match the market prices of long-term variance swaps reasonably well.

A variance swap is a financial product that allows investors to trade realized variance against current implied variance of log returns. 
More specifically, let $S_t$ stand for the closing price of the underlying asset on day $t$, $t=0, 1, \ldots, T$, and let $R_t = \log(S_t/S_{t-1})$ represent the $t$th daily log return. 
The annualized realized variance over $T$ trading days is defined as $\sigma_{\rm realized}^{2} = \frac{A}{T}\sum_{t=1}^{T} R_t^{2}$,
where $A$ is the number of trading days per year, which on average is 252.  The payoff of a variance swap is defined as
\[
N_{\rm var}(\sigma_{\rm realized}^{2} - \sigma_{\rm strike}^{2}) ,
\]
where the variance notional $N_{\rm var}$ and variance strike $\sigma_{\rm strike}^{2}$ are specified before the sale of a variance swap contract.

Variance swaps provide  investors with pure exposure to the variance of the underlying asset without directional risk. It is notably liquid across major equities, indices, and stock markets, and is growing across other markets. Historical evidence indicates selling variance systematically is profitable.

There are numerous methods in the literature of pricing variance swaps, both analytically  and numerically (see \cite{zhulian:tex} for an extensive review). Nevertheless, a pricing formula or procedure that relies on a certain stochastic process, for instance L\'{e}vy process, may suffer from a lack of parsimony, or might not fit the real data well due to the inappropriateness of model assumptions (see for instance \cite{prl:tex}).

In this section, we propose a moment-based method in conjunction with our PC risk-neutral density estimate to price variance swaps, which is free of model assumption.

Assuming the existence of a risk-neutral measure $\mathbb{Q}$, the fair price $VS_{t,T}$ of a variance swap on day $t$ is the discounted expected payoff
\begin{equation}
VS_{t,T}
=e^{-R_{tT} } N_{\rm var} \left[ \mathbb{E}_t^{\mathbb{Q}} \left( \frac{A}{T}\sum_{i=1}^{T}R_{i}^{2}
\right)- \sigma_{\rm strike}^2 \right] 
\label{eq:vsprice}
\end{equation}
where  $R_{tT}$ is the cumulative risk-free interest rate from $t$ to $T$ defined in Section~\ref{sec:nonpara}. 
To proceed, we further assume that 
\begin{assumption}\label{assume:ind}
	The increments of the process $\log{S_t}$ are independent, that is, $ \log (S_{t+1}/S_t)$ is independent of $S_0, \ldots, S_t$, $t=0, \ldots, T-1$.
\end{assumption} 
Consequently, the fair price of a variance swap can be represented by a sequence of the risk neutral moments of the underlying asset. 
\begin{proposition}\label{prop:ind}
	Assuming the existence of a risk-neutral measure $\mathbb{Q}$ and Assumption \ref{assume:ind} is fulfilled, the fair price of variance swap is
	\begin{eqnarray}
	VS_{t, T} &=& e^{-R_{tT} } N_{\rm var} \left\{\frac{A}{T} \sum_{i=1}^{t} R_i^2  +   \frac{A}{T} \mathbb{E}_t^{\mathbb{Q}}(\log S_T)^2 - \frac{A}{T} (\log S_t)^2\right.  \nonumber\\ 
	&-& \left.  \frac{2A}{T} \sum_{i=t+1}^{T}  \left[ \mathbb{E}_t^{\mathbb{Q}}\log S_{i-1}  \mathbb{E}_t^{\mathbb{Q}}\log{S_i}  -(\mathbb{E}_t^{\mathbb{Q}}\log{S_{i-1}})^2\right]    - \sigma_{\rm strike}^2 \right\}. 
	\label{eq:vspricefull}
	\end{eqnarray}
\end{proposition}
The proof of Proposition~\ref{prop:ind} is relegated to Appendix~\ref{sec:varswap}.

\subsection{Moments calculation}\label{sec:momcal}

In view of Equation~\eqref{eq:vspricefull}, pricing variance swaps requires estimating the first and second moments of $\log S_i$ under the risk-neutral measure.  One option is to use a moment-based method described by \cite{bakshi:pctex}. In this section, we employ an alternative way of calculating the moments which makes use of the proposed nonparametric approach. 

Recall that the step function $f_\Delta$ defined in \eqref{eqn:pdfpc} or \eqref{eq:fdelta} provides an approximation to the risk-neutral density $f_\Q$ of $\log S_T$. We use all the available market prices of options to estimate $f_\Delta$, then the moments calculated from $f_\Delta$ serve as the estimates of required moments. Since $f_\Delta$ is piecewise constant, it can be verified that the first and second moments of $\log(S_T)$ are given by
\begin{align}
\mathbb{E}_t^{\mathbb{Q}} \log(S_T) & = \sum_{l=1}^{q+1}  \frac{a_l}{2} [(\log K_l)^2 - (\log K_{l-1})^2], \label{eq:nonparamean}\\
\mathbb{E}_t^{\mathbb{Q}} [\log(S_T)]^2 & 
= \sum_{l=1}^{q+1} \frac{a_l}{3} [(\log  K_l)^3 - (\log K_{l-1})^3].  \label{eq:nonparamean2}
\end{align}
Note that there are no market prices available for options that expire on a day that is other than the third Saturday of the delivery month. We would have to interpolate the mean and standard deviation of $\log(S_i)$ for $t<i<T$, and this is achieved via linear interpolation in this paper. Detailed procedures are described in Appendix~\ref{sec:interpolation}.

\subsection{Replicating by variance futures}

In order to evaluate the fair price of a variance swap, we replicate variance swap using available market prices of variance futures. Variance future is a financial contract that is traded over the counter. As stated in \cite{bisc:tex}, variance swap and variance future are essentially the same since they both trade the difference of variance and one can replicate a variance swap by the corresponding variance future. As a matter of fact, if variance future and variance swap share the same expiration date, then at the start point of the observation period, there is no difference between trading a variance future and trading a variance swap with \$50 variance notional.
The formula for the fair price of a variance swap contract induced from variance future is given by
\[
VS_{t, T}=e^{-R_{tT} } N_{\rm var} \left\{   \frac{A}{T} \left[\sum_{i=1}^{M-1}R_{i}^{2} +  IUG\times \frac{N_e-M+1}{A} \times \frac{1}{100^2}\right ]- \sigma_{\rm strike}^2 \right\}, 
\]
where $M$ is the number of observed days to date, $N_e$ is the expected number of trading days in the observation period, $IUG$ is the square of market implied volatility given by
\[
IUG = \sum_{i=M}^{N_e}  R_i^2 \times \frac{A}{N_e - M+1} \times 100^2.
\]

\subsection{Variance future data}

Variance future data were downloaded from the Chicago Board Options Exchange (CBOE) website (\url{http://cfe.cboe.com/}). Variance future products with 12-month (with futures symbol VA) or 3-month (with futures symbol VT) expirations are traded on the CBOE Futures Exchange.
We use VA in the subsequent analysis. The continuously compounded zero-coupon interest rates cover dates from January 2, 1996 to August 31, 2015. For variance futures, the trading dates are from December 10, 2012 to August 31, 2015, with start dates from December 21, 2010 to July 30, 2015 and expiration dates from January 18, 2013 to January 1, 2016. We use variance futures to replicate variance swaps, so the time spans of variance swaps are in line with those of variance futures.

\subsection{Results}

In order to assess the accuracy of our estimated fair prices of a variance swap, we compare three relevant quantities:    \begin{enumerate}
	\item {\bf OP}: Fair price of a variance swap based on our moment-based nonparametric approach, using option market prices till day $t$;
	\item {\bf VF}: Induced market price of a variance swap from CBOE traded variance future till day $t$;
	\item {\bf True}: Realized price of a variance swap at expiration day $T$ with known $S_0, S_1, \ldots, S_T$.
\end{enumerate}

\begin{figure}
	\begin{subfigure}{.5\textwidth}
		\centering
		\includegraphics[height = 6cm,width=\linewidth]{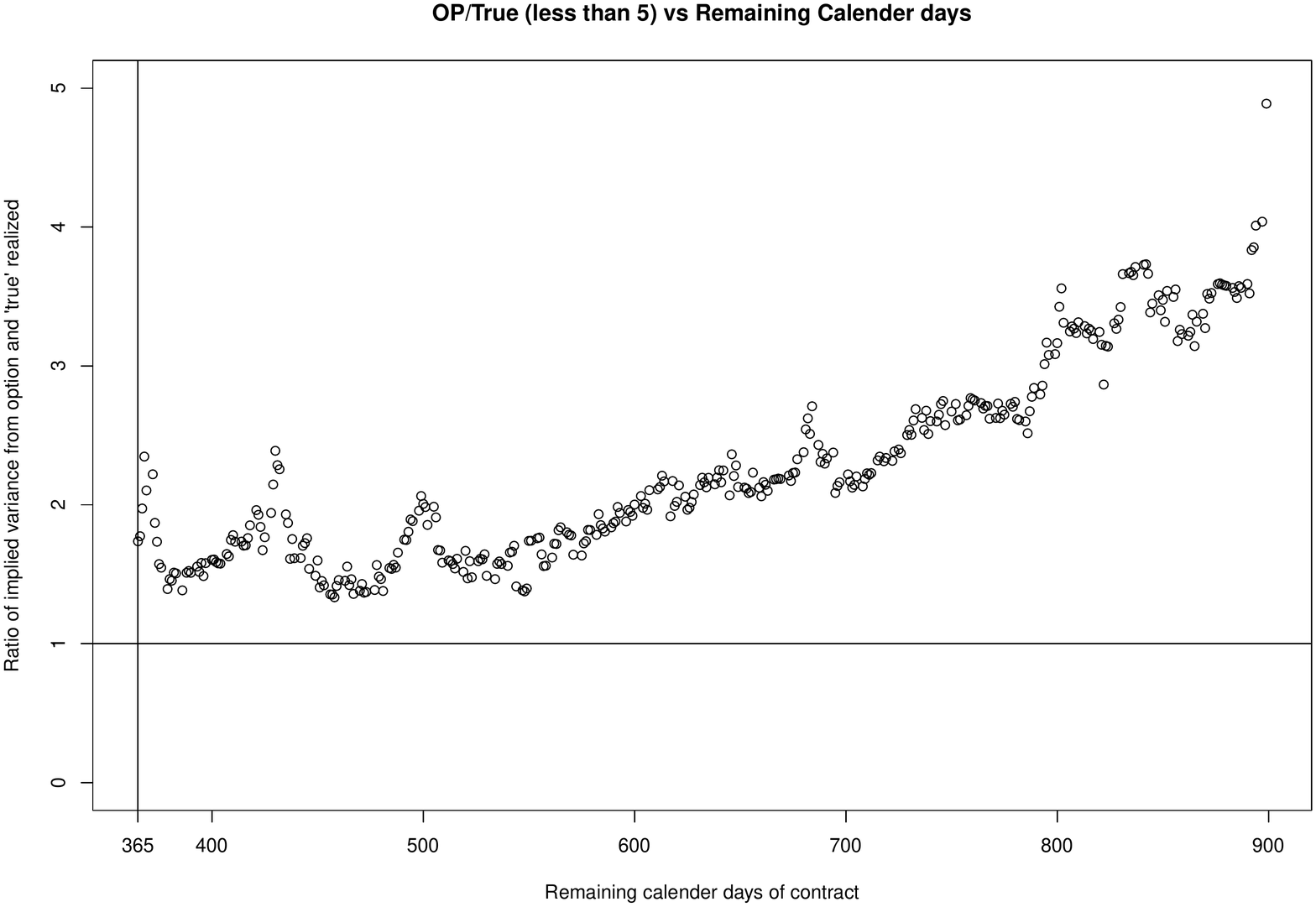}
		\caption{Ratio of OP/True vs. days to expiration}
		\label{fig:op2}
	\end{subfigure}%
	\begin{subfigure}{.5\textwidth}
		\centering
		\includegraphics[height = 6cm,width=\linewidth]{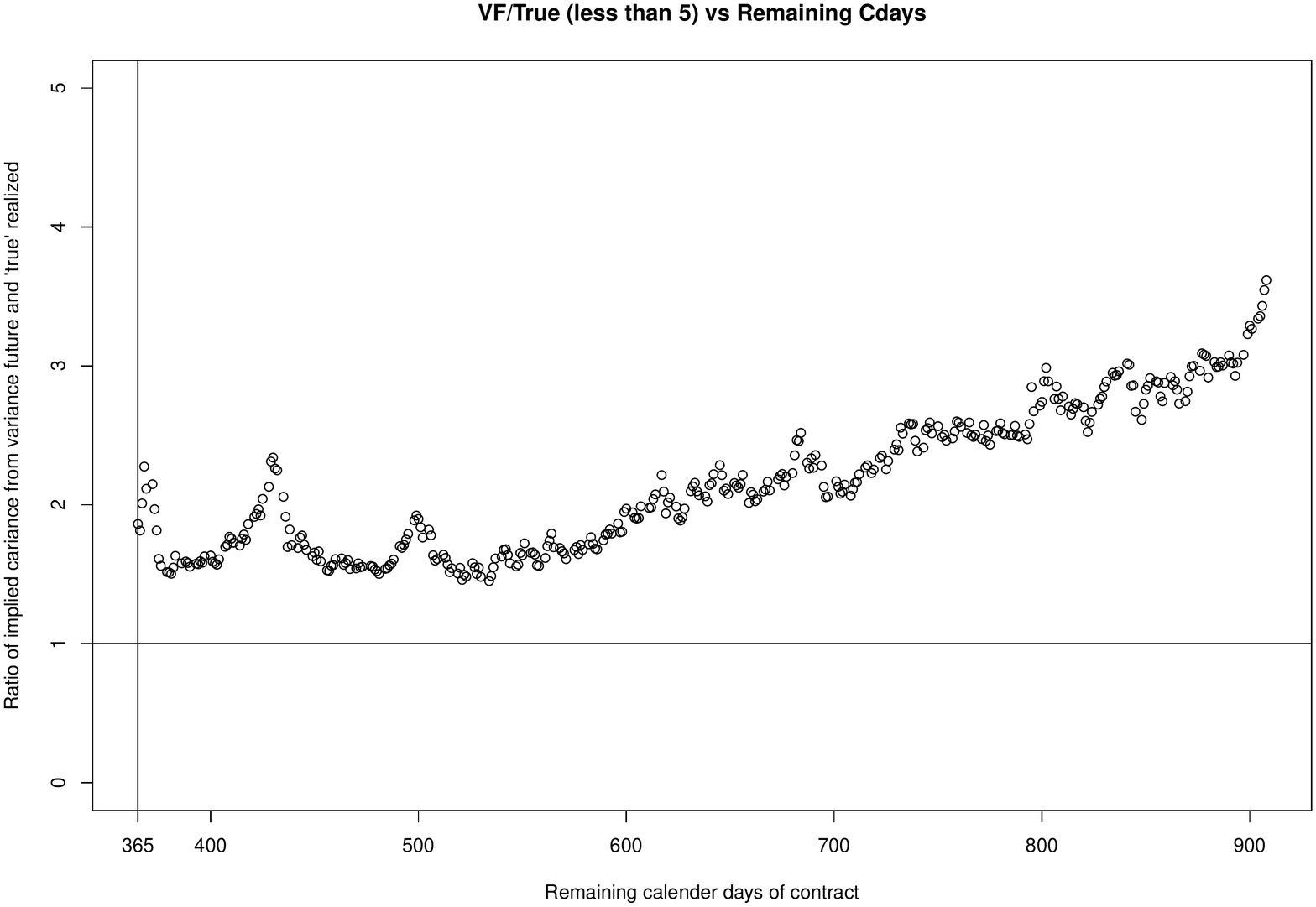}
		\caption{Ratio of VF/True vs. days to expiration}
		\label{fig:vf1}
	\end{subfigure}%
	
	\begin{subfigure}{.5\textwidth}
		\centering
		\includegraphics[height = 6cm,width=\linewidth]{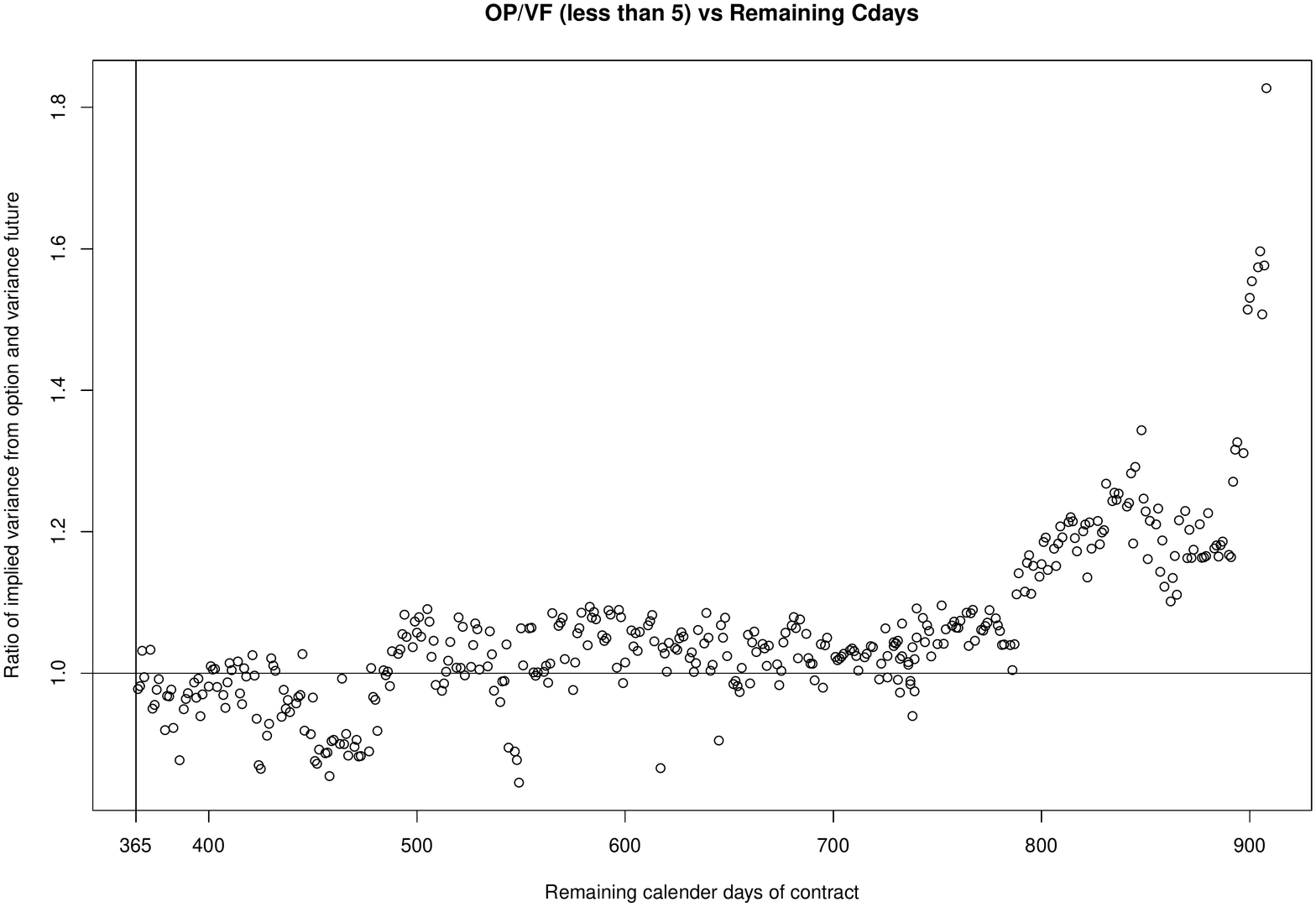}
		\caption{Ratio of OP/VF vs. days to expiration}
		\label{fig:opvf4}
	\end{subfigure}%
	\begin{subfigure}{.5\textwidth}
		\centering
		\includegraphics[height = 6cm,width=\linewidth]{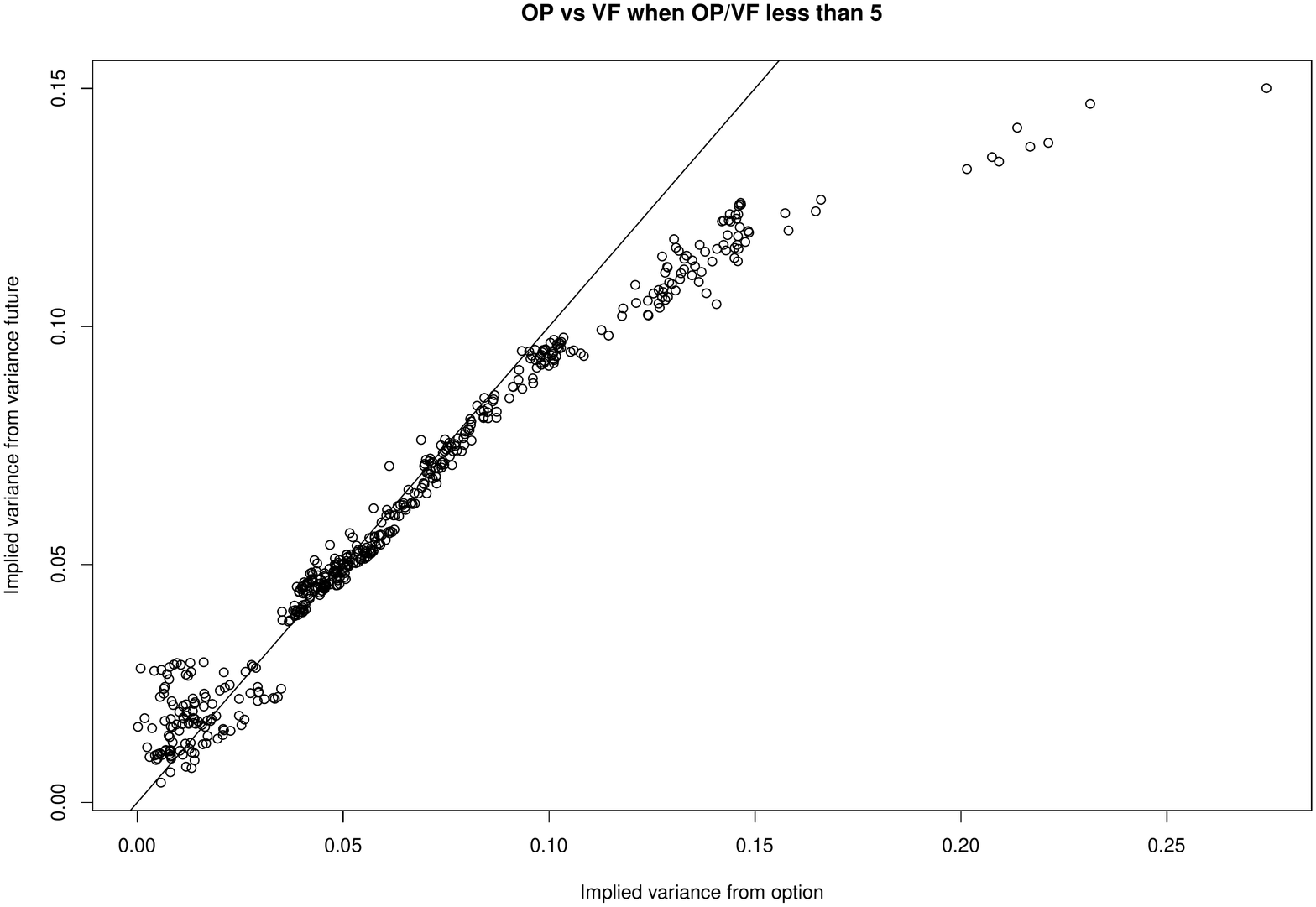}
		\caption{OP vs. VF}
		\label{fig:opvf2}
	\end{subfigure}%
	\caption{Comparison OP, VF and True variance swap prices}
	\label{fig:vsall1}
\end{figure}

We present three ratios, OP/True, VF/True, and OP/VF, in Figure~\ref{fig:vsall1}, against the remaining calendar days of variance swaps.  Compared with ``True'' prices based on realized underlying asset prices, Figure~\ref{fig:vsall1}(\subref{fig:op2}) and Figure~\ref{fig:vsall1}(\subref{fig:vf1}) suggest that OP and VF have a similar increasing pattern along with the remaining calendar days. This is in part due to the uncertainty in the estimate of the variance, which increases with the number of days to expiration.
On the other hand, Figure~\ref{fig:vsall1}(\subref{fig:opvf4}) and Figure~\ref{fig:vsall1}(\subref{fig:opvf2}) show that the fair price OP based on our proposed method matches the market price VF pretty well on variance swaps with expiration between 365 days and 800 days. For variance swaps expiring in less than 365 days (not shown here), OP and VF do not match well. This is plausibly attributed to the fact that long-term options are more reasonable and stable, which are less likely to be affected by external factors or noises. For variance swaps longer than 800 days, the relatively low VF might indicate underpriced variance futures.

\section{Discussion}\label{sec:conclude}

In this paper, we propose a new nonparametric approach for estimating the RND. It is data-driven, and is not built on any model assumption about the data generating process of underlying asset prices. It only assumes the existence of a risk-neutral density and the independence of increments of log return for pricing variance swaps. That is why it can capture the market price very well. 

In contrast with other nonparametric methods, such as cubic spline and B-spline, our method is much simpler but fit the real data better. We choose only distinct strikes as knots and assume constant values between knots to avoid  overfitting. By sacrificing the continuity of estimated risk-neutral density, the non-negativity of a density function is readily satisfied.  

On the other hand, the proposed approach utilizes market prices of all options, not just OTM options. In our opinion, ITM options, despite not being as liquid as OTM options, still contain market information and should be incorporated when estimating a risk-neutral density. Our comprehensive analysis shows that it recovers OTM option prices better by including ITM option prices. 

Pricing variance swaps is a difficult job when dealing with real data. We display in Figure~\ref{fig:vsall1} only the cases where the ratio OP/True is less than 5. There are cases where OP and VF disagree significantly. Overall, our OP prices work better for variance futures that expire in the last four months of 2015, which are also the last four months available in our dataset.

\vskip 14pt
\noindent {\large\bf Acknowledgements}

We thank Dr.~Liming Feng from the University of Illinois at Urbana-Champaign and Ms.~Yuhang Liang from Northwestern University for their extremely help during data collection. 
\par


\section*{Appendix}

\appendix

\section{Proof of Proposition~\ref{prop:callputs}}\label{sec:proofcallputs}

We rewrite the call and put option prices in Equations~\eqref{eq: estrnd1} and \eqref{eq: estrnd2} in terms of $a_1, a_2, \ldots, a_q, a_{q+1}$ as follows

\begin{equation} \label{eq:equ1app}
\begin{aligned}
e^{R_{tT}} \hat{P}_i 
&= \int_{-\infty}^{\log K_i} (K_i - e^y) f_\Delta(y) dy\\
&= \sum_{l=1}^{q+1} \int_{\log K_{l-1}}^{\log K_l} (K_i - e^y) a_l dy    \cdot  \mathds{1} (K_i \geq K_l )\\
&= \sum_{l=1}^{q+1} a_l [(K_i  \log\frac{K_l}{K_{l-1}}) - (K_l - K_{l-1})]   \cdot  \mathds{1} (K_i \geq K_l ), \ \text{$i \in \mathcal{P}$}
\end{aligned}
\end{equation}

\begin{equation}\label{eq:equ2app}
\begin{aligned}
e^{ R_{tT}}\hat{C}_{i}
&= \int_{\log K_{i}}^{\infty}  (e^y - K_{i}) f_\Delta(y) dy\\
&=  \sum_{l=1}^{q+1} \int_{\log K_{l-1}}^{\log K_{l}} (e^y - K_{i}) a_{l} dy  \cdot  \mathds{1} (K_{i} \leq K_{l-1}) \\
&= \sum_{l=1}^{q+1}   a_{l}[ (K_{l} - K_{l-1}) - K_{i} \log\frac{K_{l}}{K_{l-1}}] \cdot  \mathds{1} (K_{i} < K_{l}) ,\  \text{$i \in \mathcal{C}$}
\end{aligned}
\end{equation}

Let $X^{(p)}_{i,l} = [K_i  \log(K_l / K_{l-1}) - (K_l - K_{l-1})]   \cdot  \mathds{1} (K_i \geq K_l )$, $l = 1, 2, \ldots, q+1$ be an entry of the design matrix for put options; and $X^{(c)}_{i,l} = [ (K_{l} - K_{l-1}) - K_{i} \log (K_{l}/K_{l-1})] \cdot  \mathds{1} (K_{i} < K_{l})$, $l = 1, 2, \ldots, q+1$ for call options. 
From Equation~\eqref{eq:unityct},  $a_{q+1}$ can be represented by $a_1, a_2, \ldots, a_q$, as
\begin{equation} \label{eq:alast}
a_{q+1} = \left(1-\sum_{l=1}^{q} a_l \log\frac{K_l}{K_{l-1}}\right)(\log c_K)^{-1} 
\end{equation}
Plugging Equation~\eqref{eq:alast} into Equations~\eqref{eq:equ1app} and \eqref{eq:equ2app}, we obtain
\begin{equation} \label{eq:equ3A}
\begin{aligned}
e^{R_{tT}} \hat{P}_i &= \sum_{l=1}^{q+1} a_l X^{(p)}_{i,l}\\
&= a_1X^{(p)}_{i,1} +a_2X^{(p)}_{i,2} +\cdots +a_qX^{(p)}_{i,q}\\
&\quad +\left(1 - a_1\log\frac{K_1}{K_0} - \cdots - a_q \log\frac{K_q}{K_{q-1}}\right)(\log c_K)^{-1} X^{(p)}_{i,q+1}\\ 
&=a_1[X^{(p)}_{i,1}-(\log\frac{K_1}{K_0} )(\log c_K)^{-1} X^{(p)}_{i,q+1}]\quad+\cdots \\
&\quad +a_q[X^{(p)}_{i,q} - (\log\frac{K_q}{K_{q-1}} )(\log c_K)^{-1} X^{(p)}_{i,q+1}] + \frac{1}{\log c_K}X^{(p)}_{i,q+1}\\
&\buildrel\triangle\over =a_1 X^{(P)}_{i,1} + a_2 X^{(P)}_{i,2} + \cdots + a_q X^{(P)}_{i,q} +  X^{(P)}_{i,q+1}, \text{ $i \in \mathcal{P}$}
\end{aligned}
\end{equation}
where $X^{(P)}_{i,l}  = X^{(p)}_{i,l} -(\log K_l / K_{l-1} )(\log c_K)^{-1} X^{(p)}_{i,q+1}$, $l = 1, 2, \ldots, q$ and $X^{(P)}_{i,q+1}  = X^{(p)}_{i,q+1} / \log c_K$~.
Similarly for call options,
\begin{equation}\label{eq:equ4A}
\begin{aligned}
e^{R_{tT}} \hat{C}_{i} &= \sum_{l=1}^{q+1} a_l X^{(c)}_{i,l}\\
&= a_1X^{(c)}_{i,1} +a_2X^{(c)}_{i,2} +\cdots +a_qX^{(c)}_{i,q} \\
&\quad+ \left(1 - a_1\log\frac{K_1}{K_0} - \cdots - a_q\log\frac{K_q}{K_{q-1}}\right)(\log c_K)^{-1} X^{(c)}_{i,q+1}\\ 
&=a_1[X^{(c)}_{i,1}-(\log \frac{K_1}{K_0} )(\log c_K)^{-1} X^{(c)}_{i,q+1}] \quad+\cdots\\
&\quad+a_q[X^{(c)}_{i,q} - (\log \frac{K_q}{K_{q-1}} )(\log c_K)^{-1}  X^{(c)}_{i,q+1}] + \frac{1}{\log c_K}X^{(c)}_{i,q+1}\\
&\buildrel\triangle\over =a_1 X^{(C)}_{i,1} + a_2 X^{(C)}_{i,2} + \dots + a_q X^{(C)}_{i,q} +  X^{(C)}_{i,q+1}, \text{ $i \in \mathcal{C}$}
\end{aligned}
\end{equation}
where $X^{(C)}_{i,l} = X^{(c)}_{i,l}-(\log  K_l / K_{l-1} )(\log c_K)^{-1}X^{(c)}_{i,q+1}$, $l = 1, \ldots, q$ and $X^{(C)}_{i,q+1} = X^{(c)}_{i,q+1} /\log c_K$~.
\hfill{$\Box$}

\section{Proof of Theorem~\ref{thm:consistency}}\label{sec:theorem3.1proof}

Given $\epsilon > 0$, let $\delta_1 = \sqrt{\epsilon} e^{R_{tT}}/[3(1+c_K+e)] > 0$. There exists $-\infty < A < 0 < B < \infty$, such that, 
\[
\int_{-\infty}^A f_\Q(x) dx < \delta_1, \>
\int_{-\infty}^A e^x f_\Q(x) dx < \delta_1, \>
\int_B^{\infty} f_\Q(x) dx < \delta_1, \>
\int_B^{\infty} e^x f_\Q(x) dx < \delta_1
\]
Let $\delta_2 = \sqrt{\epsilon} e^{R_{tT}-B-1}/[3(B-A+2)] > 0$. Since $f_\Q$ is continuous, there exists a $\delta > 0$, such that, for any $x_1, x_2 \in [A-1, B+1]$, 
$$|f_\Q(x_1) - f_\Q(x_2)| < \delta_2$$
as long as $|x_1 - x_2| < \delta$.

For small enough $K_1, |\Delta|$ and large enough $q, K_q$, there exist integers $u, v$, such that, $1 < u < u+1 < v < v+1 < q$, $\log K_u \leq A < \log K_{u+1}$, $\log K_v < B \leq \log K_{v+1}$, $|\Delta| < \delta$. 

We construct a $f_\Delta$ by defining
\begin{eqnarray*}
	a_1 &=& (\log c_K)^{-1} \int_{-\infty}^{\log K_1} f_\Q(x) dx\> \geq\> 0\\
	a_i &=& [\log(K_i/K_{i-1})]^{-1} \int_{\log K_{i-1}}^{\log K_i} f_\Q(x)dx\> \geq\> 0,\>\>\> i=2, \ldots, q\\
	a_{q+1} &=& (\log c_K)^{-1} \int_{\log K_q}^{\infty} f_\Q(x) dx\> \geq \>0
\end{eqnarray*}
It can be verified that $\int_{-\infty}^{\infty} f_\Delta (x) dx = \sum_{i=1}^{q+1} a_i \log (K_i/K_{i-1}) = 1$. Let 
\[
\Delta_f = \max_{u\leq i\leq v} \left( \max_{\log K_i \leq x\leq \log K_{i+1}} f_\Q(x) - \min_{\log K_i \leq x\leq \log K_{i+1}} f_\Q(x)\right)
\]
Then $|\Delta| < \delta$ implies $\Delta_f \leq \delta_2$. It can be verified that
\begin{eqnarray*}
	|\hat{C}_i - \tilde{C}_i| &<& \left\{
	\begin{array}{cl}
		\sqrt{\epsilon}/3, & \mbox{ for }i=v+1, \ldots, q\\
		2\sqrt{\epsilon}/3, & \mbox{ for }i=u, \ldots, v\\
		\sqrt{\epsilon}, & \mbox{ for }i=1, \ldots, u-1
	\end{array}
	\right.\\
	|\hat{P}_i - \tilde{P}_i| &<& \left\{
	\begin{array}{cl}
		\sqrt{\epsilon}/3, & \mbox{ for }i=1, \ldots, u\\
		2\sqrt{\epsilon}/3, & \mbox{ for }i=u+1, \ldots, v+1\\
		\sqrt{\epsilon}, & \mbox{ for }i=v+2, \ldots, q
	\end{array}
	\right.
\end{eqnarray*}
In other words, there exist $a_1, \ldots, a_{q+1}$, such that, $(\hat{C}_i - \tilde{C}_i)^2 < \epsilon$, $(\hat{P}_i - \tilde{P}_i)^2 < \epsilon$, for $i=1, \ldots, q$. It implies the $(a_1, \ldots, a_{q+1})$ that minimizes $L(a_1, \ldots, a_{q+1})$ also satisfies
\[
\frac{1}{2q} \left[\sum_{i=1}^q (\hat{C}_i - \tilde{C}_i)^2 + \sum_{i=1}^q (\hat{P}_i - \tilde{P}_i)^2\right] < \epsilon
\]
which leads to the conclusion.
\hfill{$\Box$}

\section{Proof of Proposition~\ref{prop:ind} }\label{sec:varswap}

Since $\mathbb{E}_t^{\mathbb{Q}}[\sum_{i=1}^{T}R_{i}^{2}]  = \sum_{i=1}^{t} R_i^2  +  \sum_{i=t+1}^{T} \mathbb{E}_t^{\mathbb{Q}}[R_i^2]$, the key part 
\[
\begin{aligned}
&  \sum_{i=t+1}^{T} \mathbb{E}_t^{\mathbb{Q}}[R_i^2] 
=\sum_{i=t+1}^{T} \mathbb{E}_t^{\mathbb{Q}} [\log \frac{S_i}{S_{i-1}}]^2\\
&=\sum_{i=t+1}^{T}[\mathbb{E}_t^{\mathbb{Q}}(\log S_{i})^{2}+\mathbb{E}_t^{\mathbb{Q}}(\log S_{i-1})^{2}-2\mathbb{E}_t^{\mathbb{Q}}(\log S_{i})(\log S_{i-1})]
\\
&=\sum_{i=t+1}^{T}\mathbb{E}_t^{\mathbb{Q}}(\log S_{i})^{2}+\sum_{i=t+1}^{T}\mathbb{E}_t^{\mathbb{Q}}(\log S_{i-1})^{2}-2\sum_{i=t+1}^{T}\mathbb{E}_t^{\mathbb{Q}}[\log S_{i-1}+\log (\frac{S_i}{S_{i-1}})][\log S_{i-1}]\\
&=\sum_{i=t+1}^{T}\mathbb{E}_t^{\mathbb{Q}}(\log S_{i})^{2}+\sum_{i=t+1}^{T}\mathbb{E}_t^{\mathbb{Q}}(\log S_{i-1})^{2}-2\sum_{i=t+1}^{T}\mathbb{E}_t^{\mathbb{Q}}(\log S_{i-1})^{2}\\
&\quad -2\sum_{i=t+1}^{T}\mathbb{E}_t^{\mathbb{Q}}[\log S_{i-1}][\log (\frac{S_i}{S_{i-1}})]\\
&=\mathbb{E}_t^{\mathbb{Q}}[\log S_T]^2-[\log S_t]^2-2\sum_{i=t+1}^{T}\mathbb{E}_t^{\mathbb{Q}}[\log S_{i-1}][\log (\frac{S_i}{S_{i-1}})]\\
& = \mathbb{E}_t^{\mathbb{Q}}[\log S_T]^2-[\log S_t]^2-2\sum_{i=t+1}^{T}\mathbb{E}_t^{\mathbb{Q}}[\log S_{i-1}]\mathbb{E}_t^{\mathbb{Q}}[\log (\frac{S_i}{S_{i-1}})]\\
& =  \mathbb{E}_t^{\mathbb{Q}}[\log S_T]^2-[\log S_t]^2-2\sum_{i=t+1}^{T}  [ \mathbb{E}_t^{\mathbb{Q}}\log S_{i-1}  \mathbb{E}_t^{\mathbb{Q}}\log {S_i}  -(\mathbb{E}_t^{\mathbb{Q}}\log {S_{i-1}} )^2]  
\end{aligned}
\]
Then \eqref{eq:vspricefull} can be obtained by plugging $\mathbb{E}_t^{\mathbb{Q}}[\sum_{i=1}^{T}R_{i}^{2}]$ into \eqref{eq:vsprice}. 
\hfill{$\Box$}

\section{Linear interpolation for 1st and 2nd moments}\label{sec:interpolation}

\paragraph{Mean imputation}
Suppose the trading day is $t$ and the expiration day is $T$.  We denote all possible expiration dates of traded contracts by $t+n_1, t+n_2, \dots$. Suppose the time point to be imputed is $t+n_0$. Given all the information available at day $t$, $\log S_{t}$ can be regarded as its expectation at day $t$, $\mathbb{E}^{\mathbb{Q}}_t \log S_{t}$. Therefore, we consider cases separately according to whether or not $t+n_0$ is in the interval $[t, t+n_1]$ and then apply linear interpolation to obtain the mean of $\log S_{t+n_0}$. More specifically, there are two cases:
\begin{enumerate}
	\item \textbf{Case 1}: $n_0 \in [0, n_1]$ and $\mathbb{E}^{\mathbb{Q}}_t(\log S_{t+n_1})$ has been calculated. 
	\[
	\begin{aligned}
	\mathbb{E}^{\mathbb{Q}}_t(\log S_{t+n_0}) 
	&=\mathbb{E}^{\mathbb{Q}}_t(\log S_{t+n_1}) - \frac{ (n_1 - n_0) [\mathbb{E}^{\mathbb{Q}}_t(\log S_{t+n_1})- \log S_t]}{n_1}\\
	&=\frac{ n_0\mathbb{E}^{\mathbb{Q}}_t (\log S_{t+n_1})+(n_1-n_0)\log(S_t)}{n_1}
	\end{aligned}
	\]
	
	\item \textbf{Case 2}: $n_0 \in [n_i, n_{i+1}]$ for some $i=1, 2, \ldots $. The expectations $\mathbb{E}^{\mathbb{Q}}_t(\log S_{t+n_i})$ and $\mathbb{E}^{\mathbb{Q}}_t(\log S_{t+n_{i+1}})$ have already been calculated. 
	\[
	\begin{aligned}
	\mathbb{E}^{\mathbb{Q}}_t(\log S_{t+n_0}) 
	&= \frac{(n_0 - n_i)[\mathbb{E}^{\mathbb{Q}}_t(\log S_{t+n_{i+1}}) - \mathbb{E}^{\mathbb{Q}}_t(\log S_{t+n_i})]}{n_{i+1}- n_{i}}+ \mathbb{E}^{\mathbb{Q}}_t(\log S_{t+n_i}) \\
	&=\frac{ (n_0-n_i)\mathbb{E}^{\mathbb{Q}}_t (\log S_{t+n_{i+1}})+(n_{i+1}-n_0)\mathbb{E}^{\mathbb{Q}}_t(\log S_{t+n_i})}{n_{i+1}-n_i} 
	\end{aligned}
	\]
	
\end{enumerate}

\paragraph{Variance Imputation}

In order to calculate the variance $\mathbb{V}^{\mathbb{Q}}_t(\log S_{t+n_0})$ at day $t$, we use a similar interpolation based on the available variances of log returns at day $t$ with expiration $T$. Based on the scatterplot (not shown here) of all available variances that we have from the existing contracts, the trend of variances has a curved pattern against the number of days to expiration. More specifically, it is roughly a quadratic curve. Before we implement a linear interpolation, we first perform a square-root transformation of variances.

\begin{enumerate}
	\item \textbf{Case 1}: $n_0 \in [0, n_1]$. $\mathbb{V}^{\mathbb{Q}}_t(\log S_{t+n_1})$ has been calculated. Then
	\[
	\sqrt{\mathbb{V}^{\mathbb{Q}}_t(\log S_{t+n_0}) }
	=\frac{ n_0 \sqrt{\mathbb{V}^{\mathbb{Q}}_t (\log S_{t+n_1})}}{n_1}
	\]
	
	\item \textbf{Case 2}: $n_0 \in [n_i, n_{i+1}]$ for some $i=1, 2, \ldots $. The values $\mathbb{V}^{\mathbb{Q}}_t(\log S_{t+n_i})$ and $\mathbb{V}^{\mathbb{Q}}_t(\log S_{t+n_{i+1}})$ have been calculated. Then
	\[
	\begin{aligned}
	& \sqrt{\mathbb{V}^{\mathbb{Q}}_t(\log S_{t+n_0})}  \\
	&= \sqrt{\mathbb{V}^{\mathbb{Q}}_t(\log S_{t+n_0})} - \sqrt{\mathbb{V}^{\mathbb{Q}}_t(\log S_{t+n_i})} + \sqrt{\mathbb{V}^{\mathbb{Q}}_t(\log S_{t+n_i})} \\
	&= \frac{(n_0 - n_i)\left[ \sqrt{\mathbb{V}^{\mathbb{Q}}_t(\log S_{t+n_{i+1}})} - \sqrt{\mathbb{V}^{\mathbb{Q}}_t(\log S_{t+n_i})}\right]}{n_{i+1}- n_{i}}+ \sqrt{\mathbb{V}^{\mathbb{Q}}_t(\log S_{t+n_i})} \\
	&=\frac{ (n_0-n_i) \sqrt{\mathbb{V}^{\mathbb{Q}}_t (\log S_{t+n_{i+1}})}+(n_{i+1}-n_0) \sqrt{\mathbb{V}^{\mathbb{Q}}_t(\log S_{t+n_i})}}{n_{i+1}-n_i} .
	\end{aligned}
	\]
\end{enumerate}
Then the second moment is  
\[\mathbb{E}^{\mathbb{Q}}_t(\log S_{t+n_0})^2 = [\mathbb{E}^{\mathbb{Q}}_t(\log S_{t+n_0})]^2 + \mathbb{V}^{\mathbb{Q}}_t(\log S_{t+n_0})\] 
A fair price of variance swap $VS_{t,T}$ can be obtained by the pricing formula \eqref{eq:vsprice}.


\begin{thebibliography}{10}
	\providecommand{\url}[1]{{#1}}
	\providecommand{\urlprefix}{URL }
	\expandafter\ifx\csname urlstyle\endcsname\relax
	\providecommand{\doi}[1]{DOI~\discretionary{}{}{}#1}\else
	\providecommand{\doi}{DOI~\discretionary{}{}{}\begingroup
		\urlstyle{rm}\Url}\fi
	
	\bibitem[Bakshi et~al.(2003)]{bakshi:pctex}
	Bakshi, G., Kapadia, N., Madan, D.: Stock return characteristics, skew laws,
	and the differential pricing of individual equity options.
	\newblock The Review of Financial Studies \textbf{16}(1), 101--143 (2003)
	
	\bibitem[Biscamp and Weithers(2007)]{bisc:tex}
	Biscamp, L., Weithers, T.: Variance swaps and cboe s\&p 500 variance futures.
	\newblock Chicago Trading Company, LLC.  (2007).
	\newblock Euromoney papers
	
	\bibitem[Bliss and Panigirtzoglou(2002)]{robertnik:tex}
	Bliss, R., Panigirtzoglou, N.: Testing the stability of implied probability
	density functions.
	\newblock Journal of Banking \& Finance \textbf{26}(2-3), 381--422 (2002)
	
	\bibitem[Breeden and Litzenberger(1978)]{breedenlit:tex}
	Breeden, D., Litzenberger, R.: Prices of state-contingent claims implicit in
	option prices.
	\newblock The Journal of Business \textbf{51}(4), 621--651 (1978)
	
	\bibitem[Carr et~al.(2012)]{prl:tex}
	Carr, P., Lee, R., Wu, L.: Variance swaps on time-changed l$\acute{e}$vy
	processes.
	\newblock Finance and Stochastics \textbf{16}(2), 335--355 (2012)
	
	\bibitem[Eriksson et~al.(2004)]{eriksson2004approximating}
	Eriksson, A., Ghysels, E., Forsberg, L.: Approximating the probability
	distribution of functions of random variables: A new approach.
	\newblock Cirano (2004)
	
	\bibitem[Eriksson et~al.(2009)]{eriksson2009normal}
	Eriksson, A., Ghysels, E., Wang, F.: The normal inverse gaussian distribution
	and the pricing of derivatives.
	\newblock Journal of Derivatives \textbf{16}(3), 23 (2009)
	
	\bibitem[Ghysels and Wang(2014)]{eric:pctex}
	Ghysels, E., Wang, F.: Moment-implied densities: Properties and applications.
	\newblock Journal of Business \& Economic Statistics \textbf{32}(1), 88--111
	(2014)
	
	\bibitem[Jarrow and Rudd(1982)]{jarrowrudd:tex}
	Jarrow, R., Rudd, A.: Approximate option valuation for arbitrary stochastic
	process.
	\newblock Journal of Financial Economics \textbf{10}(3), 347--369 (1982)
	
	\bibitem[Lee(2014)]{shlee:tex}
	Lee, S.: Estimation of risk-neutral measures using quartic b-spline cumulative
	distribution functions with power tails.
	\newblock Quantitative Finance \textbf{14}(10), 1857--1879 (2014)
	
	\bibitem[Melick and Thomas(1997)]{melickthomas:tex}
	Melick, W., Thomas, C.: Recovering an asset's implied pdf from option prices:
	An application to crude oil during the gulf crisis.
	\newblock Journal of Financial and Quantitative Analysis \textbf{32}(1),
	91--115 (1997)
	
	\bibitem[Monteiro et~al.(2008)]{anarehaluis:tex}
	Monteiro, A., Tutuncu, R., Vicente, L.: Recovering risk-neutral probability
	density functions from options prices using cubic splines.
	\newblock European Journal of Operational Research \textbf{187}(2), 525--542
	(2008)
	
	\bibitem[Ritchey(1990)]{ritchey:tex}
	Ritchey, R.: Call option valuation for discrete normal mixtures.
	\newblock Journal of Financial Research \textbf{13}(4), 285--296 (1990)
	
	\bibitem[Sherrick et~al.(1992)]{sherrick1992:tex}
	Sherrick, B., Irwin, S., Forster, D.: Option-based evidence of the
	nonstationarity of expected s$\&$p 500 futures price distributions.
	\newblock Journal of Futures Markets \textbf{12}(3), 275--290 (1992)
	
	\bibitem[Zhu and Lian(2010)]{zhulian:tex}
	Zhu, S.P., Lian, G.H.: A closed-form exact solution for pricing variance swaps
	with stochastic volatility.
	\newblock Mathematical Finance \textbf{21}(2), 233--256 (2010)
	
\end{thebibliography}
\end{document}